\documentstyle[aps]{revtex}
\begin{document}
\draft
\title{Nonrelativistic Particle in Free Random Gauge Background }
\author{Michael Engelhardt\thanks{email: hengel@wicc.weizmann.ac.il} }
\address{Department of Condensed Matter Physics \\
Weizmann Institute of Science \\ Rehovot 76100, Israel }
\date{}
\maketitle

\begin{abstract}
The problem of a nonrelativistic particle with an internal color
degree of freedom, with and without spin, moving in a free random gauge
background is discussed. Freeness is a concept developed recently in
the mathematical literature connected with noncommuting random
variables. In the context of large-N hermitian matrices, it means
that the the multi-matrix model considered contains no bias with
respect to the relative orientations of the matrices. In such a
gauge background, the spectrum of a colored particle can be solved
for analytically. In three dimensions, near zero momentum, the 
energy distribution for the spinless particle displays a gap, while
the energy distribution for the particle with spin does not.
\end{abstract}


\section{Introduction}
Much of the complexity of strong interaction dynamics derives from
the non-commutativity of the gauge fields, encoded in their matrix
character. It leads to the gluonic self-interactions, which generate 
phenomena such as asymptotic freedom, and presumably also confinement
due to the constriction of chromoelectric fields into flux tubes.
Matrix-valued degrees of freedom are notoriously difficult to handle,
even in cases where one has managed to eliminate the space-time
dependence of the matrices. One such example may be the large-$N_C $
limit of QCD ($N_C $ denoting the number of colors) \cite{hoof}
\cite{cole}. There one argues that gauge field path integrals should
be dominated by a large-$N_C $ saddle point, and that there is a gauge
where this saddle point, commonly referred to as the ``master field'',
is constant in space and time. This is plausible in view of the 
translational invariance in space-time of the underlying action and 
measure.

In the case of large-$N_C $ QCD, it is not even known how to formulate
a potential purely in terms of space-time independent matrices which
will reproduce the master field; however, also if one posits some potential
of one's choice, such simplified matrix models generically become 
intractable as soon as more than one matrix degree of freedom is involved,
i.e. as soon as non-commutativity is allowed to play a role.
There are notable exceptions, such as the Itzykson-Zuber integral
\cite{itzz}, which lies at the heart of the Kazakov-Migdal and
Penner models \cite{kazm} \cite{nwei}. In these models,
due to the special type of potential involved, some selected observables
can be calculated.

Recently, there has been some advance in the mathematical literature
concerning a type of non-commuting variables called free random
variables \cite{voic}; it has furthermore been shown (loc.cit.) that in the 
large-N limit, independent hermitian matrix models describe variables
of this type. These are models governed by a partition function of the
form
\begin{equation}
Z = \int \prod_{i=1}^{D} [dA_i ]
\exp \left( -N \sum_{i=1}^{D} \mbox{Tr} \, V(A_i ) \right)
\end{equation}
where the
integration is to be carried out over the real and imaginary parts of the
matrix elements of the $A_i $, subject to the constraints of hermiticity.
The potential, consisting of a sum of terms each only involving one
of the matrices, thus only determines the eigenvalue distributions of
the $A_i $ but there is no bias with respect to the relative ``orientations''
of the matrix degrees of freedom; each orientation is equally probable.
This restriction is quite strong; most interesting matrix models,
presumably including the one describing the master field of QCD, contain
interactions between the different matrices. It should be noted, though,
that the constructions used in the theory of free random variables may be
extended to more general models and have recently led to some increased
understanding of the functional analytic spaces on which the master
field may be formally defined \cite{doug} \cite{ggross} \cite{aref}.

However, even noninteracting matrix models constitute interesting
laboratories for the study of phenomena associated with non-commutativity.
Initially, it is well known how to derive the eigenvalue distributions
of the involved matrices from their respective potentials. The eigenvalue
distribution is related by a Hilbert transform to the derivative of the
potential; i.e. if the hermitian matrix $A$ is governed by a potential
term Tr$\, V(A)$, then the eigenvalue distribution (normalized to unity)
is determined by
\begin{equation}
\frac{1}{2} V^{\prime } (z) = \int
d\lambda \, \rho (\lambda ) \frac{ {\cal P} }{z-\lambda }
\end{equation}
Then, however, as soon as one attempts to calculate eigenvalue
distributions of objects composite in the different matrices contained
in the model, one is faced with their non-commutative nature. Here, the
methods developed in \cite{voic} exhibit their full power; they enable
one to convolute the distributions of different free random variables
in a quite general and systematic manner, not restricted to a very
special form of the objects considered.

In this paper, two not altogether trivial cases of problems which can be
solved completely analytically with the techniques developed in 
\cite{voic} are exhibited. These are the problems of
nonrelativistic particles with an internal color degree of freedom,
with or without spin, moving in a free Gaussian random gauge background.
The objective of this work is to develop some intuition regarding the
phenomenological consequences of freeness in the context of large-N
matrix degrees of freedom. This can be viewed as complementary to the
recent more formal investigations mentioned above concerned with the
formal definition of master fields and the spaces on which they
operate. Here, by contrast, the emphasis lies with the study of some
exactly solvable model systems which may serve as paradigms for more
complicated realistic cases. The examples treated here, while
sufficiently simple to be of pedagogical value, nevertheless seem
generic enough to be relevant for applications. Some
physics questions for which they may be of relevance will be mentioned
at the end of the paper.

It is to be expected that developing some insight into the phenomena
resulting from freeness also is of some value from the point of view of the
quest to understand interacting matrix models beyond the few existing
solvable cases. With the advent of techniques allowing one to deal with
free variables, one controls the two limiting cases of vanishing and
infinitely strong coupling between matrix degrees of freedom: A system
of matrices whose orientations are completely aligned due to some
strong coupling is as simple as a one-matrix model, since all matrices
can be simultaneously diagonalised; on the other hand, a system of
matrices whose relative orientations are subject to no bias at all
can now be treated as well. Knowing both limits should provide valuable
qualitative predictions for the behaviour of models even with nontrivial
couplings.

\section{Particle without Spin}
Consider the Hamiltonian
\begin{equation}
H_0 = \sum_{i=1}^{D} (k_i -A_i )^2
\label{hs0}
\end{equation}
This is the Hamiltonian of a spinless nonrelativistic particle with
color in $D$ dimensions, minimally coupled to a background vector
field $\vec{A} $. Each of the components of $\vec{A} $ is taken to be an
$N\times N$ hermitian matrix (where $N\rightarrow \infty $), constant in
space and time. Due to this last condition, the Hamiltonian is diagonal
in momentum space and the solutions can be classified according to the
momentum components $k_i $. For given $\vec{k} $ and $\vec{A} $, there
are $N$ different modes of propagation, i.e. color eigenvectors, for
the particle. Now, consider the vector field not as fixed, but to be
averaged over an ensemble described by an independent matrix model.
In other words, any observable $O$ should be averaged as
\begin{equation}
\langle O \rangle = \frac{1}{Z} \int \prod_{i=1}^{D} [dA_i ] O(\vec{A} )
\exp \left( -N \sum_{i=1}^{D} \mbox{Tr} \, V(A_i ) \right)
\end{equation}
Note that now the model is invariant under arbitrary unitary
rotations of the variables $A_i $.

In short, the problem proposed is the following: Given the eigenvalue
distributions of the $A_i $ ( which can be easily found from the 
potential $V(A_i ) $, as mentioned in the introduction), find the
eigenvalue distribution of $H_0 $ for arbitrary $k_i $. Since the model
contains no bias with respect to the relative orientations of the $A_i $,
the $A_i $ constitute a family of free random variables, and the
problem is a fairly straightforward application of the techniques
developed in \cite{voic}. This will serve as a warmup for the more
involved case of the spinning particle, where one will have to deal
with more exotic situations, such as the free random variables themselves
appearing as components of matrices.

To treat a concrete model, the potential $V(A_i )$ must be specified.
If one wishes to preserve invariance of the model under spatial
rotations without destroying the freeness, one is forced to choose
a Gaussian potential, $V(A)= \beta A^2 $. The eigenvalue distribution for
the matrices $A_i $ induced by this potential is the well-known
semicircular distribution \cite{mehta}\footnote{In the following,
a factor $N$ will be taken out of eigenvalue distributions, i.e.
$\rho_{A} $ is normalized to unity.},
\begin{equation}
\rho_{A} (\lambda ) = \frac{2}{\pi a^2 } \sqrt{a^2 - \lambda^{2} }
\label{semic}
\end{equation}
where $a^2 = 2/\beta $.

As a first step in finding the eigenvalue distribution of the
Hamiltonian, one easily finds the eigenvalue distribution of each of
the summands $(k_i -A_i )^2 $ of $H_0 $: If one has 
$\rho_{A} (\lambda ) d\lambda $ eigenvalues of $A_i $ in the interval
$[\lambda , \lambda + d\lambda ]$, then the number of eigenvalues of
$(k_i - A_i )^2 $ to be found in the interval
$[(k_i - \lambda )^2 , (k_i -\lambda )^2 -2(k_i - \lambda ) d\lambda ] $
is $(\rho_{A} (\lambda ) + \rho_{A} (-\lambda ))\, d\lambda $.
Substituting $\epsilon = (k_i -\lambda )^2 $, one thus finds
$\rho_{(k_i -A_i )^2 } (\epsilon ) d\epsilon $ eigenvalues of
$(k_i -A_i )^2 $ in the interval $[\epsilon , \epsilon + d\epsilon ]$,
where
\begin{equation}
\rho_{(k_i -A_i )^2 } (\epsilon ) = \frac{1}{2\sqrt{\epsilon } }
(\rho_{A} (\sqrt{\epsilon } + k_i ) + \rho_{A} (\sqrt{\epsilon } -k_i ) )
\label{indd}
\end{equation}
Of course now there are only positive eigenvalues, $\epsilon > 0$.

It remains to convolute the eigenvalue distributions of the different
summands $(k_i -A_i )^2 $. These are hermitian random variables, with
eigenvalue distributions given by (\ref{indd}), and freely rotating with
respect to one another, i.e. there is no bias in the relative orientations
in color space of the variables when taking the ensemble average over
the $A_i $. To convolute these eigenvalue distributions, one must compute
their so-called R-transform \cite{voic}\footnote{This problem has 
recently also been treated from a more physical point of view in
\cite{azee}.}. The R-transform plays the same
role for adding noncommuting free random variables as the logarithm of the
Fourier transform does for adding ordinary commuting random variables:
Convolution becomes ordinary addition, i.e. adding the R-transforms of
the eigenvalue distributions of the summands gives the R-transform of the
eigenvalue distribution of the sum. The R-transform is defined by the
following algorithm: First, find the generating function of the moments
of the distribution:
\begin{eqnarray}
G(\xi ) &=& \frac{1}{\xi } + \sum_{n=1}^{\infty } \rho_{n} 
\frac{1}{\xi^{n+1 } } \\
\rho_{n} &=& \int_{-\infty }^{\infty } d\lambda \, 
\rho (\lambda ) \lambda^{n}
\end{eqnarray}
which can be analytically continued to the entire upper complex half-plane
by summing the geometric series:
\begin{equation}
G(\xi ) = \int_{-\infty }^{\infty } d\lambda \,
\frac{\rho (\lambda ) }{\xi -\lambda }
\label{ancong}
\end{equation}
This continuation is important for the inverse problem of going from
the generating function back to the distribution \cite{tamar},
\begin{equation}
\rho (\lambda ) = -\frac{1}{\pi } \lim_{\eta \searrow 0 } \mbox{Im} \,
G(\lambda + i\eta )
\label{stie}
\end{equation}
Next, find the function inverse to $G$, i.e. $K$ such that
$K(G(\xi )) = \xi $. Then, the R-transform is given by
$R(z) = K(z)-1/z $.

It turns out that this program can still be carried out analytically for
the program at hand; one is led to algebraic equations of at most fourth
order and thus just remains within the realm of solvability. The
generating function corresponding to the distribution (\ref{indd})
is easily calculated to be
\begin{eqnarray}
G_i (\xi ) &=& \int_{0}^{\infty } d\epsilon \, \frac{1}{\xi -\epsilon }
\frac{\rho_{A} (\sqrt{\epsilon } + k_i ) + \rho_{A} (\sqrt{\epsilon }
-k_i )}{2\sqrt{\epsilon } } \\
&=& \int_{-\infty }^{\infty } d\epsilon \,
\frac{\rho_{A} (\epsilon + k_i )}{\xi - \epsilon^{2} } \\
&=& \frac{2}{\pi } \int_{-1}^{1} d\epsilon \,
\frac{\sqrt{1-\epsilon^{2} } }{(\sqrt{\xi } + k_i -a\epsilon )
(\sqrt{\xi } - k_i + a\epsilon )} \\
&=& \frac{1}{a^2 } \left( 2-\frac{1}{\sqrt{\xi } } (
\sqrt{(\sqrt{\xi } - k_i )^2 -a^2 } +
\sqrt{(\sqrt{\xi } + k_i )^2 -a^2 } ) \right)
\label{invgi}
\end{eqnarray}
Now, to find the inverse $K_i $ of $G_i $, one must solve (\ref{invgi})
for $\xi $ in terms of $G_i $.
By always bringing the terms containing no square roots to
the left hand side and squaring the resulting equation (this procedure must
be applied twice), one arrives at an equation with no square roots;
in this equation, quadratic in $\xi $, the constant term vanishes, the
trivial solution $\xi =0 $ is spurious, and thus one arrives at
\begin{equation}
R_i (z) +\frac{1}{z} \equiv
K_i (z) \equiv \xi (z) = \frac{1}{z} + \frac{a^2 }{4-a^2 z}
+\frac{4k_i^2 }{(2-a^2 z)^2 }
\label{indr}
\end{equation}
The R-transform of the eigenvalue distribution of the Hamiltonian is now
simply the sum of the individual R-transforms corresponding to each
spatial component:
\begin{equation}
R_{H_0 } (z) = \sum_{i=1}^{D} R_i (z) = \frac{Da^2 }{4-a^2 z}
+\frac{4k^2 }{(2-a^2 z)^2 }
\label{rtra}
\end{equation}
where $k^2 $, the squared Euclidean length of the vector $\vec{k} $,
has been introduced. Note that the $k_i $ dependence of the individual
R-transforms $R_i $ could have already been predicted from the 
additivity of the R-transform and the rotational invariance of the
Hamiltonian\footnote{Remember that the original Gaussian potential
governing the distribution of the vector field components is invariant
under rotation of the vector field in the spatial indices.}. From
$\sum_{i=1}^{D} f(k_i ) = F (|\vec{k} |) $ it already follows that
$f$ must be of the form $f(k_i ) = c+d k_i^2 $, as is manifest in
(\ref{indr}).

It remains to go from the R-transform $R_{H_0 } $ back to the eigenvalue 
distribution of the Hamiltonian. This is the more arduous task. Inverting
$K_{H_0 } (z) \equiv R_{H_0 } (z) + 1/z$ corresponds to solving the equation
\begin{equation}
4(2-a^2 z)^2 + (D-1)a^2 z (2-a^2 z)^2 + 4k^2 z (4-a^2 z)
=\xi z (4-a^2 z) (2-a^2 z)^2
\end{equation}
for $z$ in terms of $\xi $. This is a fourth order equation for $z$,
which via the substitution
\begin{equation}
z=x-\frac{D-1}{4\xi } + \frac{2}{a^2 }
\label{subs}
\end{equation}
can be brought to the normal form
\begin{equation}
x^4 + (3N-6M^2 ) x^2 + (8M^3 - 6NM) x + 3NM^2 - 3M^4 + L/4 = 0
\label{red4}
\end{equation}
where the abbreviations
\begin{equation}
L= \frac{64 k^2 }{a^8 \xi } \ \ \ \ \ \ \ \ M=\frac{D-1}{4\xi } 
\ \ \ \ \ \ \ \ N=\frac{4}{3a^2 \xi } \left( 1+\frac{D-1}{2}
-\frac{k^2 }{a^2 } \right)
-\frac{4}{3a^4 }
\label{pard}
\end{equation}
have been introduced. The analysis of such an equation is quite
involved, and for this reason is relegated to the appendix.
The final result for the eigenvalue distribution of the Hamiltonian
(\ref{hs0}), which according to (\ref{stie}) is essentially the imaginary
part of the solution of (\ref{red4}), can be stated as follows:
\begin{equation}
\rho_{H_0 } (\xi ) =
\frac{1}{2\pi } \theta (-d) (1-\theta (-b_2 ) \theta (b_1 ) )
\left| \sqrt{ | y_2 | } + (1-2\theta (4M^3 -3NM )) \sqrt{ | y_3 | } \right|
\label{fuls}
\end{equation}
where
\begin{eqnarray}
y_2 = 2\mbox{Re} \, (u e^{2\pi i/3} ) -2N + 4M^2 & \ \ \ \ \ \ \ \ &
y_3 = 2\mbox{Re} \, (u e^{-2\pi i/3} ) -2N + 4M^2 \\
u=\left( -\frac{q}{2} + i\sqrt{-d} \right)^{1/3} & &
d=\frac{p^3 }{27} + \frac{q^2 }{4} \\
p= -3N^2 -L & & q= -2N^3 + 2NL - 4M^2 L \\
b_2 = 6N-12M^2 & &
b_1 = 9N^2 - 48NM^2 + 48M^4 - L
\end{eqnarray}
The solution $\rho_{\sqrt{H_0 } } (\lambda ) = 
2\lambda \rho_{H_0 } (\lambda^{2} )$ is plotted (this representation
avoids the trivial square root singularity at zero eigenvalue, especially
further below in the case of the spinning particle) for various
values of the free parameters $k/a$ and $D$ in figures 
(\ref{f1})-(\ref{f3}). Note that when comparing different $D$,
there are two distinct sensible ways of choosing the units.
On the one hand, one may take the width $a$ of the distributions of the
vector fields as the unit of the square root of energy, as one does when
comparing different $k$ (cf. figures (\ref{f1}) and (\ref{f2})). 
However, one may also rescale $a^2 = \tilde{a}^{2} /D$ and use
$\tilde{a}$ as the unit of the square root of energy. This corresponds to
keeping the norm of the vector field, i.e. $\sum_{i=1}^{D} \mbox{Tr} \,
A_i^2 $, constant as one varies $D$ (just as one automatically
does in the case of the momentum $\vec{k} $ by only talking 
about its modulus), cf. figure (\ref{f3}).

To characterize the behaviour of the solution more precisely, it
is useful to evaluate various moments and limits. The first few
moments of the distribution $\rho_{H_0 } $ can be obtained without
knowledge of the full solution (\ref{fuls}). It is sufficient to
expand the R-transform (\ref{rtra}) in a power series in $z$ and invert
the corresponding series $K_{H_0 } (z) \equiv R_{H_0 } (z) + 1/z$ to the
desired order, yielding the first few terms of $G(\xi ) $ expanded
in powers of $1/\xi $. The moments are then by definition the
coefficients appearing in $G(\xi ) $, yielding
\begin{eqnarray}
\left\langle \frac{1}{N} \mbox{Tr} \, H_0 \right\rangle &=& 
\frac{a^2 D}{4} + k^2 = \frac{\tilde{a}^{2} }{4} +k^2 \\
\left\langle \frac{1}{N} \mbox{Tr} \, H_0^2 \right\rangle - 
\left\langle \frac{1}{N} \mbox{Tr} \, H_0 \right\rangle^{2} 
&=& \frac{a^4 D}{16} + a^2 k^2
= \frac{\tilde{a}^{4} }{16 D} + \frac{\tilde{a}^{2} k^2}{D}
\end{eqnarray}
With the help of this information, one can obtain the behaviour
of the distribution as $k\rightarrow \infty $ or $D\rightarrow \infty $.

Consider large $D$, and use $a^2 $ as the unit of energy.
The distribution is centered around $a^2 D/4 $,
with a width behaving as $\sqrt{D} $. Substituting therefore
\begin{equation}
\xi = \frac{a^2 D}{4} + \gamma \sqrt{D}
\end{equation}
in (\ref{fuls}), and keeping only leading pieces in $1/\sqrt{D} $, one
obtains that $b_2 < 0 $ always, and $b_1 < 0 $ for
$\gamma \in [-a\sqrt{k^2 + a^2 }/2 , a\sqrt{k^2 +a^2 }/2]$. Also,
\begin{equation}
d=\frac{2^{18} (4\gamma^{2} -a^4 ) k^2 }{27a^{30} } \, \frac{1}{D^2 }
\end{equation}
i.e. $d<0$ for $\gamma \in [-a^2 /2 , a^2 /2 ]$. The condition on
$d$ is more stringent than the one on $b_1 $ and thus defines the
support of the distribution. Furthermore, one has
\begin{eqnarray}
M^2 -3N/4 &=& -\frac{4}{a^8 } (a^2 (a^2 -k^2 ) -4\gamma^{2} ) \frac{1}{D} \\
| y_2 | &=& \frac{16}{a^8 } \left( a^2 (a^2 + k^2 ) -4\gamma^{2}
+2ak\sqrt{a^4 -4\gamma^{2} } \right) \frac{1}{D} \\
| y_3 | &=& \frac{16}{a^8 } \left( a^2 (a^2 + k^2 ) -4\gamma^{2}
-2ak\sqrt{a^4 -4\gamma^{2} } \right) \frac{1}{D} \\
\sqrt{|y_2 |} \pm \sqrt{|y_3 |} &=&
\sqrt{|y_2 | + |y_3 | \pm 2\sqrt{|y_2 | \, |y_3| } } \\
&=& \frac{4\sqrt{2} }{a^4 }
\sqrt{ a^2 (a^2 + k^2 ) -4\gamma^{2} \pm
|a^2 (a^2 -k^2 ) -4\gamma^{2} | } \frac{1}{\sqrt{D} }
\end{eqnarray}
and therefore
\begin{equation}
\rho_{H_0 } (\xi ) = \frac{4}{\pi a^4 \sqrt{D} } \sqrt{a^4 -4\gamma^{2} }
\end{equation}
In other words, for $D\rightarrow \infty $ with fixed $k$ and $a$
one obtains a semicircular distribution centered at $a^2 D/4 $ and
with radius $a^2 \sqrt{D} /2$.

On the other hand, if one chooses to use $\tilde{a}^{2} $ as the unit
of energy, the limiting behaviour as $D\rightarrow \infty $ can already
be read off from the first two moments. The width of the distribution
vanishes and one obtains a $\delta $-peak at $\tilde{a}^{2} /4 +k^2 $.

Finally, taking the limit $k\rightarrow \infty $ can be done in complete
analogy to the limit $D\rightarrow \infty $ above. However, it is
not necessary to explicitely go through this: For $k\rightarrow \infty $,
one can approximate the Hamiltonian by $H_0 \approx k^2 + 2\sum_{i=1}^{D}
k_i A_i $. This means that one simply has to additively convolute
the original semicircle distributions of the $A_i $, with the radii
scaled by a factor $2k_i $, respectively. It is well known \cite{voic}
that the semicircle distributions are closed under additive free
convolution and that the radii add like components of an Euclidean
vector. Therefore one immediately obtains the result that, for
$k\rightarrow \infty $, the eigenvalue distribution of $H_0 $ approaches
a semicircle centered at $k^2 $ and with radius $2ka$.

The physical interpretation of this result is clear: For $k\rightarrow
\infty $, the quantum mechanical propagation of the particle is
dominated by the semiclassical straight trajectories in the direction
of $\vec{k} $. The particle does not see the transverse directions
and behaves just as it would in a one-dimensional world; correspondingly,
the dimension $D$ does not enter the limiting semicircular
eigenvalue distribution as $k\rightarrow \infty $.

On the other hand, when $k$ is finite, the particle does explore the
dimensions available to it to some extent. For large $D$,
the width behaves as $a^2 \sqrt{D} = \tilde{a}^{2} /\sqrt{D} $ (and
thus does not tend to a finite value when one scales the
vector field to have constant norm as $D$ is varied).
The semicircular nature of the
distribution in the limit of large $D$ is understandable as
a manifestation of the central limit theorem: For a large number
of variables (matrices), the sum of the random matrices (after
separating off the first moments) behaves
as if it was distributed according to a Gaussian weight,
i.e. as a semicircle \cite{voic}.

Most interesting is the soft region around
$k=0$. For $k=0$, the solution (\ref{fuls}) simplifies considerably.
In this case, one has
\begin{eqnarray}
d &=& 0 \\
b_2 &=& 6N-12M^2 \\
b_1 &=& (3N-4M^2 )(3N-12M^2 )
\end{eqnarray}
and therefore only a nonvanishing eigenvalue distribution when
$N>4M^2 /3 $. In this case, one has $\sqrt{|y_2 |} = \sqrt{|y_3 |}
=\sqrt{3N-4M^2 } $ and therefore
\begin{equation}
\rho_{H_0 } (\xi ) = \frac{1}{\pi } \theta (3N-4M^2 ) \sqrt{3N-4M^2 }
\end{equation}
The edges of the distribution are determined by
\begin{equation}
0=3N-4M^2 = -\frac{(D-1)^2 }{4\xi^{2} } + \frac{2(D+1)}{a^2 \xi } 
-\frac{4}{a^4 }
\end{equation}
leading to $\xi = a^2 (\sqrt{D} \pm 1)^2 /4$, i.e. the support of the
distribution is the interval 
$[a^2 (\sqrt{D} -1)^2 /4 , a^2 (\sqrt{D} +1)^2 /4]$. Thus, in two or
more dimensions, the eigenvalue distribution has a gap, even for 
$k\rightarrow 0$, as already evidenced in figures (\ref{f1})-(\ref{f3}). 
Of course, for $D=1$ one just has for $H_0 $ the distribution (\ref{indd}),
which behaves proportionally to $1/\sqrt{\xi } $ as $\xi \rightarrow 0$.
Note also that the above result on the support of the distribution
implies that, at $k=0$, the spectrum of the square root of $H_0 $ 
always has a support of length $a$, for any dimension.

Formally, the emergence of an energy gap does not seem particularly
surprising in view of the fact that the Hamiltonian (\ref{hs0}) is a
sum of squares. Consider the more familiar case of ordinary (commuting)
random variables, taking positive values, and with a finite probability
density when one approaches zero. Typically, the distribution of a sum
of two such variables will vanish linearly at zero, because sampling
the two variables, it is improbable to simultaneously find two very
small values. There is actually a matrix realisation of this, namely
two random matrices with the aforementioned type of distributions for
the eigenvalues, but not freely rotating over the whole $U(N)$ group
with respect to one another; instead, constrained to commute, i.e.
only all permutations of the eigenvalues are allowed. Presumably to
realise this, one would have to introduce a strong interaction term
between the two matrices, proportional to the square of their
commutator.

In the present case, by contrast, one is allowing a much larger class
of relative orientations between the matrices, namely an arbitrary
$U(N)$ rotation. This larger invariance class evidently enhances the
effect of making the occurrence of a small eigenvalue improbable in the
sum of the matrices. For free random variables, the effect is so strong
that the resulting distribution not only vanishes as one approaches zero,
but vanishes on a whole interval between zero and a finite value.

\section{Particle with Spin}
The Hamiltonian of a particle with spin in the same gauge background as
was considered in the case of the spinless particle is the $2N \times 2N$
matrix
\begin{equation}
H_S =(\vec{\sigma } (\vec{k} - \vec{A} ))^2 = (\vec{k} -\vec{A} )^2
+i\vec{\sigma } (\vec{A} \times \vec{A} )
\label{pham}
\end{equation}
where the $\sigma_{i} $ denote the Pauli matrices; the
number of space dimensions has now been specialized to $D=3$.
In the last term on the right hand side, one recognizes the coupling
of the spin to the nonabelian magnetic field; the latter contains no
derivative term since the vector potential is taken to be spatially
constant.

In practice, it is not helpful here to separate the Hamiltonian into
the spinless part and the spin-magnetic-field coupling. The crucial
question when applying the free random variable techniques is whether
one can cast the calculation into successive multiplications and additions
of mutually free variables. E.g., given two free random variables $X$
and $Y$, it is in general not straightforward to calculate the eigenvalue
distribution of $X^2 +XY$; one may accomplish the multiplicative
convolution in the second summand (how this is done is explained below), 
but the two summands do not rotate independently of one another and therefore
the free convolution techniques do not apply. On the other hand, the
eigenvalue distribution e.g. of $X+XY$ is straightforward to obtain once
one writes $X+XY = X(1+Y)$. Thus, in the case of the Hamiltonian 
(\ref{pham}), having solved the spinless problem is of no assistance;
one must restart and consider the eigenvalue distribution of the object
\begin{equation}
P=\vec{\sigma } (\vec{k} - \vec{A} )
\label{psqrh}
\end{equation}
(obtaining the distribution of $H_S =P^2 $ at the end is trivial).

As in the case of the spinless particle, the eigenvalue distributions of
the $A_i $ are taken to be semicircular with radius $a$; the harmonic
potential generating these distributions is invariant under spatial
rotations. Therefore, without loss of generality, one may take $\vec{k} $
in the 3-direction, $\vec{k} = (0,0,k)$. Being able to do this will be
crucial for the developments to follow.

One is faced here with a new situation: Free random variables appearing
as components of a matrix. Considering for the moment matrices of
finite rank, the eigenvalues $\lambda $ of $P$ for a specific realisation
of the gauge matrices $A_i $ are determined by the equation
\begin{equation}
0=\det (P-\lambda ) = \det Q_{\lambda }
\label{evdet}
\end{equation}
with the $N\times N$ matrix
\begin{equation}
Q_{\lambda } = A_3 -k-\lambda + (A_1 + iA_2 ) (A_3 -k+\lambda )^{-1}
(A_1 - iA_2 )
\end{equation}
Note that $Q_{\lambda } $ has only half the rank of $P-\lambda $; however,
since $Q_{\lambda } $ is nonlinear in $\lambda $, it can have the same
number of singular points in $\lambda $ as $P-\lambda $, as required.
The second equality in (\ref{evdet}) is only valid as long as 
$A_3 -k+\lambda $ has no exact zero eigenvalues. However, this happens
only on a set of $\lambda $ of zero measure; it suffices here to
consider the generic case where the above equality is valid. Also,
the same consideration for $\tilde{P} =(i\sigma_{2} ) P (-i\sigma_{2} ) $
shows that the ensemble averaged
eigenvalue distribution of $P$ is symmetric about zero
if the eigenvalue distributions of the $A_i $ are. Therefore, it will be
enough to consider positive $\lambda $.

When one is faced with matrix equations nonlinear
in a parameter such as $\lambda $ in the present case, it is
customary to go to an equation in larger matrices linear in the
parameter \cite{goh}. This is also familiar in the context of
``supersymmetry tricks''. Here, it is not helpful; it just leads
from $Q_{\lambda } $ back to the original matrix $P$. Instead, here
the extended problem
\begin{equation}
\det (Q_{\lambda } -\mu ) =0
\label{ext}
\end{equation}
will be considered, and at the end specialized to $\mu =0$. In the large-$N$,
ensemble averaged language, the eigenvalue distribution of $Q_{\lambda } $,
parametrically depending on $\lambda $, will be obtained using free
convolution techniques. This yields a distribution $\rho_{\lambda } (\mu )$
which describes how many solutions of (\ref{ext}) one finds at a point
in the $\lambda - \mu $ plane if one counts them as they occur on an
increment $d\mu $ in $\mu $-direction. In the end, this will have to
be translated into how many solutions one finds if one counts them as
they occur on an increment $d\lambda $ in $\lambda $-direction; then,
one may set $\mu =0$.

One further issue must be settled before the free convolution
techniques can be applied: In general, the two summands $Q_1 $ and
$Q_2 $ appearing in $Q_{\lambda } = Q_1 +Q_2 $,
\begin{eqnarray}
Q_1 &=& A_3 -k-\lambda \label{q1def} \\
Q_2 &=& (A_1 +iA_2 ) (A_3 -k+\lambda )^{-1} (A_1 -iA_2 )
\end{eqnarray}
do not rotate independently of one another, since they both contain
$A_3 $. This would thwart attempts to use free convolution. The special
case considered here, however, can be rescued: The matrix $A_1 +iA_2 $,
with $A_1 $ and $A_2 $ hermitian, represents a general complex matrix,
which can be reparametrized using the polar decomposition \cite{hua}
\begin{equation}
A_1 + iA_2 = V^{\dagger } B V U^{\dagger }
\label{specd}
\end{equation}
where $U$ and $V$ are unitary and $B$ is diagonal and positive.
Integrating over all hermitian $A_1 $ and $A_2 $ corresponds to
integrating over $U$ and $V$ with Haar measure and over the
eigenvalues $\eta $ contained in $B$,
\begin{equation}
[dA_1 ][dA_2 ] = J(\vec{\eta } ) d^N \eta \, [dU][dV]
\end{equation}
where the Jacobian is 
\begin{equation}
J(\vec{\eta } ) = \prod_{i=1}^{N} \eta_{i} \prod_{i<j} (\eta_{i}^{2}
-\eta_{j}^{2} )^2
\label{poldj}
\end{equation}
To be precise, the integration over $V$ is only over the quotient
group $U(N)/(U(1)^N)$; this is no consequence, as will be explained
below. Also, note that in (\ref{specd}), the product $VU^{\dagger } $ is
usually written as a single matrix; however, since $U$ is being integrated
over with Haar measure, one can always pull out a factor $V$.

In the new variables, the potential governing $A_1 $ and $A_2 $
becomes
\begin{equation}
\mbox{Tr} \, V(A_1 ,A_2 ) = \beta \mbox{Tr} \, (A_1^2 + A_2^2 )
=\beta \mbox{Tr} \, B^2
\label{dpot}
\end{equation}
and one has
\begin{equation}
Q_2 = V^{\dagger } B V U^{\dagger } (A_3 -k+\lambda )^{-1} U
V^{\dagger } B V
\end{equation}
Cast in this form, it becomes clear that the calculation of the
$Q_{\lambda } $ eigenvalue distribution can be carried out using free
convolution: The combination $U^{\dagger } (A_3 -k+\lambda )^{-1} U$
rotates independently of $Q_1 = A_3 -k-\lambda $, and therefore the
evaluation of the eigenvalue distribution of $Q_{\lambda } $ indeed
becomes a sequence of free convolutions if one additionally uses that the
eigenvalue distribution of $Q_2 $ is the same as the one of
\begin{equation}
\tilde{Q}_{2} = (V^{\dagger } BV)^{-1} Q_2 (V^{\dagger } BV)
=U^{\dagger } (A_3 -k+\lambda )^{-1} U V^{\dagger } B^2 V
\end{equation}
(up to regions of zero measure in $B$). It is also clear that the
restricted integration domain of $V$ is of no consequence: For two
variables to be free with respect to one another, it suffices to
rotate one of the variables. The problem is in fact invariant under
the additional rotations by $V$ and the $V$-integration gives only
a normalization factor.

Finally, before applying multiplicative free convolution to obtain
the eigenvalue distribution of $\tilde{Q}_{2} $, it remains to solve
the saddle point equation for the eigenvalue distribution of $B^2 $,
which is controlled by the Jacobian (\ref{poldj}) and the potential
(\ref{dpot}). This standard procedure is carried out in the 
Appendix, yielding
\begin{equation}
\rho_{B^2 } (\eta ) = \frac{1}{a^2 \pi } \sqrt{\frac{2a^2 -\eta }{\eta } } ,
\ \ \ \ \ \ \ 0\le \eta \le 2a^2
\label{b2dist}
\end{equation}
Note that $B$ itself is therefore distributed according to a 
quarter-circle of radius $\sqrt{2} a$.

As explained in connection with the spinless particle, additive free
convolution of eigenvalue distributions is accomplished by going to the 
corresponding R-transforms, which are additive; similarly, for
multiplicative free convolution one defines the S-transform \cite{voic},
which behaves multiplicatively. The S-transform is defined as follows:
Find again the moment generating function $G(z)$ as in (\ref{ancong});
solve for the function inverse to $G(1/z)/z -1$, i.e. $T(y)$ such
that
\begin{equation}
\frac{1}{T(y)} G\left( \frac{1}{T(y)} \right) -1 = y
\end{equation}
Then the S-transform is given by
\begin{equation}
S(y)=\frac{1+y}{y} T(y)
\end{equation}
With the help of these techniques, one can now carry out the necessary
convolutions leading to the eigenvalue distribution of $Q_{\lambda } $.
Starting with the eigenvalue distributions (\ref{b2dist}) and
(cf. (\ref{semic}),(\ref{indd}))
\begin{equation}
\rho_{(A_3 -k+\lambda )^{-1} } (\eta ) = \frac{1}{\eta^{2} }
\rho_{A} \left(\frac{1}{\eta } +k-\lambda \right)
\end{equation}
one obtains the corresponding generating functions
\begin{eqnarray}
G_{B^2 } (z) &=& \frac{1}{a^2 } \left( 1-\sqrt{\frac{z-2a^2 }{z} } \right) \\
G_{(A_3 -k+\lambda )^{-1} } (z) &=&
\int_{-\infty }^{\infty } dx \frac{\rho_{A} (x) }{z-\frac{1}{x-k+\lambda } }
\\ &=& \frac{2}{\pi a^2 z} \int_{-a}^{a} dx \sqrt{a^2 -x^2 }
\left( 1+\frac{1}{z} \, \frac{1}{x-k+\lambda-\frac{1}{z} } \right) \\
&=& \frac{1}{z} - \frac{2}{z^2 a^2 } \left( \frac{1}{z} +k-\lambda
-\sqrt{(\frac{1}{z} +k-\lambda )^2 -a^2 } \right)
\end{eqnarray}
In both cases, $T$ is determined by a quadratic equation after
one has applied a procedure analogous to the one leading to 
eq. (\ref{indr}). In the case of $T_{B^2 } $, the correct solution
is easily picked out because there is a spurious zero solution;
in the case of $T_{(A_3 -k+\lambda )^{-1} } $,
the correct solution can be picked out by
considering the limiting behaviour as $a\rightarrow 0$ and using that
the S-transform of the unit matrix multiplied by a constant $w$ is 
$1/w$. Thus one arrives at the S-transforms
\begin{eqnarray}
S_{B^2 } (y) &=& \frac{2}{a^2 } \, \frac{1}{1+y} \\
S_{(A_3 -k+\lambda )^{-1} } (y) &=& \frac{\lambda -k}{2} \left( 1
+ \sqrt{1-\frac{a^2 (1+y)}{(\lambda -k)^2 } } \right) \\
S_{Q_2 } (y) = S_{\tilde{Q}_{2} } (y) =
S_{B^2 } (y) \cdot S_{(A_3 -k+\lambda )^{-1} } (y)
&=& \frac{\lambda -k}{a^2 (1+y)} \left( 1+\sqrt{1-
\frac{a^2 (1+y)}{(\lambda -k)^2 } } \right)
\end{eqnarray}
From this, one obtains that the function
\begin{equation}
\frac{G_{Q_2 } (1/z)}{z} -1 \equiv T_{Q_2 }^{-1} (z) \equiv u
\end{equation}
satisfies the equation
\begin{equation}
z^4 a^4 (1+u)^3 -2za^2 (\lambda -k) u(1+u) + a^2 u^2 = 0
\end{equation}
and consequently, $G_{Q_2 } (z) \equiv v$ satisfies
\begin{equation}
a^4 v^3 z -2a^2 (\lambda -k) v(vz-1) +a^2 (vz-1)^2 = 0
\label{gq2eq}
\end{equation}
There is no need to solve for $G_{Q_2 } $ at this point; the eigenvalue
distribution of $Q_2 $ must still be additively convoluted with the one
of $Q_1 $. Therefore, one can solve (\ref{gq2eq}) directly for
$z=G_{Q_2 }^{-1} (v)$,
\begin{equation}
K_{Q_2 } (v) \equiv G_{Q_2 }^{-1} (v) = \frac{1}{v} - \frac{a^2 v}{2}
+\lambda -k - (\lambda -k) \sqrt{ \left( 1-\frac{a^2 v}{2(\lambda -k)}
\right)^{2} - \frac{a^2 }{(\lambda -k)^2 } }
\end{equation}
where the relevant solution has been picked out by using, in the limit
$a\rightarrow 0$, that $Q_2 \rightarrow 0$ and therefore the R-transform
$R_{Q_2 } (v) = K_{Q_2 } (v) -1/v \rightarrow 0$. Combining this with
the R-transform corresponding to $Q_1 $ (cf. (\ref{q1def})),
\begin{equation}
R_{Q_1 } (v) = \frac{a^2 }{4} v -k-\lambda
\end{equation}
one finally obtains
\begin{equation}
G_{Q_{\lambda } }^{-1} (v) \equiv K_{Q_{\lambda } } (v) =
\frac{1}{v} - \frac{a^2 }{4} v -2k
-(\lambda -k) \sqrt{ \left( 1-\frac{a^2 v}{2(\lambda -k)}
\right)^{2} - \frac{a^2 }{(\lambda -k)^2 } }
\label{gqlm1}
\end{equation}
This determines $G_{Q_{\lambda } } $ and thus ultimately
$\rho_{\lambda } (\mu ) = -\mbox{Im} \, G_{Q_{\lambda } } (\mu ) / \pi $.
However, before embarking on the arduous task of inverting (\ref{gqlm1}),
it is advantageous at this stage to consider how $\rho_{\lambda } (\mu ) $
in the end determines the eigenvalue distribution $\rho_{P} (\lambda ) $
of $P$ (cf. (\ref{psqrh})). Considering for the moment finite matrices,
the solutions of equation (\ref{ext}) for a specific realization of the
gauge matrices $A_i $ define continuous trajectories in the
$\lambda -\mu $ plane, cf. figure (\ref{traj}).
Asymptotically, there are trajectories in the vicinity of the
lines defined by $\lambda +k = -\mu $ and $\lambda -k =0$, where 
``vicinity'' means at a finite distance of roughly up to the semicircle
radius $a$. The quantity one is ultimately interested in is the number
of trajectories one crosses if one marches from $\lambda =\infty $
in along the $\lambda $-axis to some point $\lambda_{0} $; this is
essentially the integral over $\rho_{P} (\lambda )$. Consider the conjecture
that this is the same as the number of trajectories one crosses
marching from $(\mu ,\lambda ) = (\infty ,\infty )$ in $\lambda $-direction
to $(\mu ,\lambda ) = (\infty ,\lambda_{0} )$, and then in $\mu $-direction
to $(\mu ,\lambda ) = (0,\lambda_{0} )$, cf. figure (\ref{traj}).
There is one point which must
be clarified before this statement can be accepted as true: In general,
it might happen that a trajectory intersects the integration paths described
above more than once, and then the two paths may count a different number
of trajectory crossings, as displayed in figure (\ref{traj}) in the
right-hand graph. It will now be argued that this is not possible. 
To begin with, note that
\begin{equation}
Q_{\lambda } (k) -\mu = Q_{\lambda +\mu /2} (k+\mu /2)
\label{mleqv}
\end{equation}
This means that solving the extended problem (\ref{ext}) with general
$\mu $ is in fact equivalent to solving the original problem (\ref{evdet})
with shifted $k$ and $\lambda $ (conversely, $k$ could be absorbed into
$\lambda $ and $\mu $ ). Now, the solutions $\lambda $ of the original
problem det$(P-\lambda )=0$ are continuous in $k$, which can be seen
as follows: Since the determinant of a Hermitian matrix is real, the
characteristic polynomial det$(P-\lambda )$ is real-valued for real
$\lambda $; its coefficients are real polynomials in $k$.
Therefore, the graph of det$(P-\lambda )$ varies continuously with $k$.
Now, if one wanted to ``annihilate'' two zeros of det$(P-\lambda )$ as
$k$ is varied, say by having a minimum of det$(P-\lambda )$ cross
from below the $\lambda $-axis to above the $\lambda $-axis, on would have
to ``create'' two other zeros somewhere else on the $\lambda $-axis in a
similar fashion, because det$(P-\lambda )$ must always have $2N$ zeros
($2N$ denoting the rank of $P$). However, at the point in $k$ where this
discontinuity takes place, one would thus have more than $2N$ zeros,
which is impossible. Therefore, the eigenvalues
are indeed continuous as $k$ is varied. According to (\ref{mleqv}), this
immediately also implies that in the extended problem 
det$(Q_{\lambda } -\mu )=0$, as $\mu $ is varied, the trajectories
$\lambda (\mu ) $ must be continuous.

Consider now the possibility that a trajectory in the $\lambda -\mu $ plane
crosses a line of constant $\mu $ more than once, as in figure (\ref{traj})
in the right-hand graph.
This would imply a discontinuous dependence $\lambda (\mu )$ and can
therefore not arise. In conclusion, this shows that marching along any
line parallel to the $\lambda $-axis in the $\lambda -\mu $ plane,
one can cross any eigenvalue trajectory at most once, as in figure
(\ref{traj}) in the left-hand graph.

One could argue in a similar way for paths in the $\mu $-direction, except
for having to be more careful due to the poles in $Q_{\lambda } $.
There is an easier way to handle paths in the $\mu $-direction once
one has treated the case of the $\lambda $-direction: Since one crosses
each trajectory at most once when integrating in $\lambda $-direction,
it is necessary to cross all of them to saturate the normalisation
condition in $\lambda $-direction. Then, however, one must also cross all
of them if one marches in straight lines between the points
$(\mu ,\lambda) = (\infty ,\infty ) \rightarrow (\infty ,\lambda_{0} )
\rightarrow (-\infty ,\lambda_{0} ) \rightarrow (-\infty ,-\infty )$.
In other words, in this way one counts at least $2N$ trajectories, and more
if there are multiple intersections when marching in $\mu $-direction.
However, one can easily check at the end using the large-$N$, 
ensemble averaged result that the integral is exactly 2 (remember that
the factor $N$ has been scaled out everywhere in the eigenvalue
distributions) and therefore realisations of the $A_i $ which 
generate multiple intersections when marching in $\mu $-direction
give vanishing contributions to the large-$N$ eigenvalue distributions.

In essence, the basic property which makes these arguments work is that
the trajectories in the $\lambda -\mu $ plane can be interpreted in
two different ways. Obviously, when considering the extended problem
(\ref{ext}), one is considering the behaviour of the eigenvalues $\mu $
under changes of the parameter $\lambda $. On the other hand, due to
the equivalence (\ref{mleqv}), one can also interpret the trajectories
as describing the behaviour of the eigenvalues $\lambda $ under changes
of the parameter $\mu $ (up to the trivial additional linear shift in 
$\lambda $). These two interpretations taken together ensure the
monotonicity of the trajectories, i.e. that they can cross each line
of constant $\lambda $ or $\mu $ only once.

As a result of these observations one can now indeed relate
$\rho_{\lambda } (\mu ) $ to $\rho_{P} (\lambda ) $:
\begin{equation}
\int_{\lambda_{0} }^{\infty } d\lambda \, \rho_{P} (\lambda ) =
\int_{0}^{\infty } d\mu \, \rho_{\lambda_{0} } (\mu ) +
\int_{\lambda_{0} }^{\infty } d\lambda \, \rho_{extra} (\lambda )
\end{equation}
which by differentiation with respect to $\lambda_{0} $ gives $\rho_{P} $.
The second contribution on the right hand side arises as follows:
For $|\lambda -k| < a$, there is a nonvanishing density of trajectories
as $\mu \rightarrow \infty $ due to the poles in $Q_{\lambda } $ (cf.
figure (\ref{traj})).
These trajectories asymptotically become parallel to the $\mu $-axis,
and therefore $\rho_{\lambda } (\mu ) $ vanishes (as it must due to
normalisation) as $\mu \rightarrow \infty $, since 
$\rho_{\lambda } (\mu ) d\mu $ measures the number of trajectories
encountered when marching in $\mu $-direction. On the other hand,
marching in $\lambda $-direction, one crosses these trajectories 
perpendicularly. For a concrete realization of the $A_i $, they are
determined as follows: In the complete problem det$(Q_1 +Q_2 -\mu )=0$,
one can neglect $Q_1 $ for $\mu \rightarrow \infty $, i.e. one is
looking for the infinite eigenvalues of $Q_2 $; these occur exactly
when $A_3 -k+\lambda $ has zero eigenvalue, i.e. when $-\lambda $
is an eigenvalue of $A_3 -k$. Therefore $\rho_{extra} (\lambda )$
is just the (shifted) original semicircle distribution,
\begin{equation}
\rho_{extra} (\lambda ) = \rho_{A} (k-\lambda )
\end{equation}
One can therefore now specify $\rho_{P} (\lambda )$:
\begin{eqnarray}
\rho_{P} (\lambda ) &=& -\frac{\partial }{\partial \lambda }
\int_{0}^{\infty } d\mu \, \rho_{\lambda } (\mu ) + \rho_{A} (k-\lambda )
\\ &=&
\frac{1}{\pi } \mbox{Im} \, \frac{\partial }{\partial \lambda }
\int_{0}^{\infty } d\mu \, G_{Q_{\lambda } } (\mu )
+ \rho_{A} (k-\lambda ) \\ &=&
\frac{1}{\pi } \mbox{Im} \, \frac{\partial }{\partial \lambda }
\int_{G_{Q_{\lambda } } (0)}^{0} dx\, x \frac{\partial }{\partial x}
G_{Q_{\lambda } }^{-1} (x) + \rho_{A} (k-\lambda ) \\ &=&
-\frac{1}{\pi } \mbox{Im} \, \frac{\partial }{\partial \lambda }
\int_{G_{Q_{\lambda } } (0) }^{0} dx \, G_{Q_{\lambda } }^{-1} (x)
+\rho_{A} (k-\lambda ) \\ &=&
-\frac{1}{\pi } \mbox{Im} \, \int_{G_{Q_{\lambda } } (0) }^{0} dx \,
\frac{\partial }{\partial \lambda } G_{Q_{\lambda } }^{-1} (x)
+\rho_{A} (k-\lambda ) \\ &=&
\frac{2}{a^2 \pi } \mbox{Im} \, \left[ (\lambda -k)
\sqrt{\left( 1-\frac{a^2 G_{Q_{\lambda } } (0) }{2(\lambda -k)} \right)^{2}
-\frac{a^2 }{(\lambda -k)^2 } } -\sqrt{(\lambda -k)^2 -a^2 } \right]
+\rho_{A} (k-\lambda )
\label{rpla}
\end{eqnarray}
(in this derivation it has been used several times that 
Im$\, G_{Q_{\lambda } } (0)$ never diverges, which will be verified below).
The last two terms in (\ref{rpla}) cancel, and using
$0=G_{Q_{\lambda } }^{-1} (G_{Q_{\lambda } } (0))$ together with
(\ref{gqlm1}), one can additionally eliminate the square root to
arrive at
\begin{equation}
\rho_{P} (\lambda ) = \frac{2}{a^2 \pi } \mbox{Im} \,
\left( \frac{1}{G_{Q_{\lambda } } (0)} -\frac{a^2 }{4} 
G_{Q_{\lambda } } (0) \right)
= -\frac{2}{a^2 \pi } \mbox{Im} \, G_{Q_{\lambda } } (0)
\left( \frac{1}{(\mbox{Re} \, G_{Q_{\lambda } } (0))^2
+(\mbox{Im} \, G_{Q_{\lambda } } (0))^2 } + \frac{a^2 }{4} \right)
\label{reimsp}
\end{equation}
It only remains thus to solve (\ref{gqlm1}) for $G_{Q_{\lambda } } (0)$.
For $v\equiv G_{Q_{\lambda } } (0)$ in (\ref{gqlm1}), the left hand
side vanishes and the square root can be eliminated by squaring the
equation. Eliminating the cubic term in the resulting quartic equation
by the substitution
\begin{equation}
v=\frac{x}{a} + \frac{4\lambda }{3a^2 }
\end{equation}
one arrives at
\begin{equation}
x^4 + c_2 x^2 + c_1 x + c_0 = 0
\label{spin4}
\end{equation}
with
\begin{eqnarray}
c_2 &=& -\frac{8}{3} (2L^2 + 4KL + 6K^2 + 1) \\
c_1 &=& -\frac{64}{27} (2L^3 + 12KL^2 + 18K^2 L + 3L -9K ) \\
c_0 &=& -\frac{16}{27} (32KL^3 + 48K^2 L^2 +8L^2 - 48KL +9)
\end{eqnarray}
given in terms of the dimensionless $K=k/a$ and $L=\lambda /a$.
The analysis of this equation is relegated to the appendix; the final
result for $G_{Q_{\lambda } } (0)$ can be stated as follows:
\begin{eqnarray}
\mbox{Im} \, G_{Q_{\lambda } } (0) &=&
-\frac{1}{2a} \theta (d) | i (\sqrt{y_1 } -\sqrt{y_2 } ) | \label{posp} \\
\mbox{Re} \, G_{Q_{\lambda } } (0) &=& \frac{1}{2a} (2\theta
(9K - 2L^3 - 12KL^2 - 18K^2 L - 3L ) -1) \sqrt{y_3 }
+\frac{4L}{3a} \ \ \ \mbox{for} \ d>0
\end{eqnarray}
where
\begin{eqnarray}
y_1 = e^{2\pi i/3} u + e^{-2\pi i/3} u^{\prime } -b_2 /3 & &
y_2 = e^{-2\pi i/3} u + e^{2\pi i/3} u^{\prime } -b_2 /3 \\
y_3 = u+u^{\prime } -b_2 /3 & \ \ \ \ \ \ \ \ &
b_2 = -\frac{16}{3} (1+2K^2 +L^2 + (2K+L)^2 ) \\
u= \frac{p}{|p|} \left| -\frac{q}{2} + \sqrt{d} \right|^{1/3} & &
u^{\prime } = -\frac{|p|}{3} \left| -\frac{q}{2} + \sqrt{d} \right|^{-1/3}
\end{eqnarray}
and
\begin{eqnarray}
d &=& \frac{p^3 }{27} + \frac{q^2 }{4} \label{disol} \\
p &=& \frac{256}{27} (2-3K^2 -9K^4 -14KL -12K^3 L +L^2 +2K^2 L^2
+4KL^3 -L^4 ) \\
q &=& \frac{4096}{729} (7-36K^2 +27K^4 +54K^6 +48KL +144K^3 L +108K^5 L
+12L^2 \nonumber \\ & & \ \ \ \ \ \ \ \ \ \
+66K^2 L^2 +18K^4 L^2 -48KL^3 -56K^3 L^3 +3L^4 -6K^2 L^4
+12KL^5 -2L^6 )
\end{eqnarray}
The solution $\rho_{\sqrt{H_S } } (\lambda ) = 2\lambda \rho_{H_S }
(\lambda^{2} ) = 2\rho_{P} (\lambda )|_{\lambda > 0 } $
is plotted for various values of the free parameter $K=k/a$ in figure
(\ref{ff1}). In the limit $k\rightarrow \infty $, the coupling of
the spin to the magnetic field in the Hamiltonian (\ref{pham}) can be
neglected and the energy spectrum is therefore identical to the spinless
case: Semicircular with radius $2ka$, centered at $k^2 $.
On the other hand, in the soft region around $k=0$, the two
cases differ qualitatively: There is no gap in the spectrum at $k=0$
for the particle with spin, since the discriminant (\ref{disol}) is
positive for $k=\lambda =0$, giving a nonzero result for (\ref{posp}).
Only at a certain nonzero value of $k$ does the distribution detach
from the origin.

In order to characterize this spin-induced effect more precisely, it
is useful to evaluate the first two moments of the energy distribution.
Since in the present case, the full generating function $G_{H_S } $
corresponding to the distribution $\rho_{H_S } $ is not explicitly
available, one cannot directly read off the moments. However, they
are easily calculated using the axioms of freeness \cite{voic}, which
the random matrices $A_i $ obey. First, one trivially has that the
first moments of the eigenvalue distributions of $H_0 $ and $H_S $ 
are the same,
\begin{equation}
\left\langle \frac{1}{2N} \mbox{Tr} \, H_S \right\rangle =
\left\langle \frac{1}{N} \mbox{Tr} \, H_0 \right\rangle = 
k^2 +\frac{3}{4} a^2
\end{equation}
since the additional spin-magnetic field coupling in $H_S $ obeys
Tr$\, (i\vec{\sigma } (\vec{A} \times \vec{A} )) =0$ already in every
fixed realization of the vector potential $\vec{A} $. Then, one has
\begin{equation}
\left\langle \frac{1}{2N} \mbox{Tr} \, H_S^2 \right\rangle -
\left\langle \frac{1}{2N} \mbox{Tr} \, H_S \right\rangle^{2} =
\left\langle \frac{1}{N} \mbox{Tr} \, H_0^2 \right\rangle -
\left\langle \frac{1}{N} \mbox{Tr} \, H_0 \right\rangle^{2} -
\left\langle \frac{1}{2N} \mbox{Tr} \, (\vec{\sigma }
(\vec{A} \times \vec{A} ) \vec{\sigma } (\vec{A} \times \vec{A} ) )
\right\rangle
\end{equation}
(again having used the tracelessness of the $\sigma $-matrices). Now,
\begin{equation}
\left\langle \frac{1}{2N} \mbox{Tr} \, (\vec{\sigma }
(\vec{A} \times \vec{A} ) \vec{\sigma } (\vec{A} \times \vec{A} ) )
\right\rangle =
\left\langle \frac{6}{N} \mbox{Tr} \, (A_1 A_2 A_1 A_2 - A_1^2 A_2^2 )
\right\rangle =
-6 \left( \frac{a^2 }{4} \right)^{2}
\end{equation}
Here, in the first equality it has been used that the matrices $A_i $
obey identical distributions; in the second equality, the axioms of
freeness for $A_1 $ and $A_2 $ have been utilized along with
$\langle (1/N) \mbox{Tr} \, A_i \rangle =0$ and
$\langle (1/N) \mbox{Tr} \, A_i^2 \rangle = a^2 /4 $. Therefore,
\begin{equation}
\left\langle \frac{1}{2N} \mbox{Tr} \, H_S^2 \right\rangle -
\left\langle \frac{1}{2N} \mbox{Tr} \, H_S \right\rangle^{2} =
\left\langle \frac{1}{N} \mbox{Tr} \, H_0^2 \right\rangle -
\left\langle \frac{1}{N} \mbox{Tr} \, H_0 \right\rangle^{2} +
\frac{3}{8} a^4 = k^2 a^2 + \frac{9}{16} a^4
\end{equation}
Thus, the effect of the additional interaction of the spin with
the magnetic field is merely to broaden, but not to shift, the
eigenvalue distribution of the Hamiltonian. It does this sufficiently
strongly to make the energy gap arising in the case of a spinless particle 
disappear at small momenta $k$. Figure (\ref{fc1}) 
compares\footnote{A factor two is divided out of $\rho_{\sqrt{H_S } } $
in order to compare two distributions normalised to unity.}
$\rho_{\sqrt{H_0 } } $ and $\rho_{\sqrt{H_S } } /2$ for different $k$.

Evidently, it is possible for the particle to align its spin with the
nonabelian magnetic field in ways which allow it to lower the energy
associated with some of the modes of propagation to values near zero.
There are sufficiently many such possibilities to make the gap in the
energy distribution disappear for small momenta $k$. A calculation of
more specific spin-color magnetic field correlation functions, which
would give a more detailed description of this effect, will be foregone
here.

\section{Summary}
In this work, the spectrum of a nonrelativistic particle with internal
colour degree of freedom moving in a constant Gaussian free random 
gauge background was discussed. Both the case of a spinless particle
and the case of a spin-1/2 particle were considered. The limit of
large momenta $k$ is easily understood perturbatively; the energy
distribution becomes semicircular. Also, in the spinless case for
a large number of spatial dimensions, one obtains a semicircle as a
consequence of the central limit theorem. On the other hand, the
nonperturbative regime near $k=0$ displays very interesting features.
The spectrum of the spinless particle exhibits a finite gap in two
or more dimensions. Evidently, for free random variables, the suppression
of small eigenvalues in sums of squares (such as the spinless Hamiltonian)
is stronger than in the case of ordinary commuting random variables,
for which only a linear vanishing of the distribution at zero occurs.
By contrast, when the particle possesses a spin degree of freedom, the
gap in the energy distribution disappears at small momenta. Formally,
this is understandable since the Hamiltonian ceases to be a sum of squares;
physically, the particle can align its spin with the background color
magnetic field such as to lower the energies of some of the modes.
This effect is sufficiently strong to obliterate the gap in the energy
spectrum occurring in the spinless case.

On a technical level, the treatment depended crucially on the fact that
the Hamiltonians considered were block-diagonal in momentum space.
Thus, it would be very hard e.g. to combine a gauge background such
as the one used here with a spatial harmonic oscillator potential for
the particle. On the other hand, one can easily envision introducing
such a gauge background into a bag model. In such a framework, the
gauge background would provide a mechanism for generating masses for
the constituent quarks. It would be interesting to pinpoint the
differences between a bag model with a random gauge background and
a conventional bag model with constituent quark masses introduced
by hand.

A related issue is the question of chiral symmetry breaking in a free
random gauge background. The chiral condensate can be related via the
Casher-Banks formula \cite{cashb} to the spectrum of the Dirac
operator \cite{verbaa} \cite{verbaaj} \cite{weid}. The methods
developed here for the $2\times 2$ spin structure occurring in the
Pauli Hamiltonian would seem to constitute a first step towards also
evaluating the eigenvalue distribution of the $4\times 4$ Dirac
operator in a random gauge background.

Finally, a different physical way to view the average over an ensemble of
spatially constant gauge configurations is to identify the ensemble
averaging with a domain or time averaging. If the gauge background
through which the particle is propagating consists of domains of
approximately constant color magnetic fields with random orientations,
then the wave packet will feel different realisations of the gauge
matrices as it evolves. Averaging over such a background can be replaced
by averaging over one domain with all possible background gauge
configurations if one argues the effects of the domain walls
to be small.

\section*{Acknowledgements}
The author acknowledges useful discussions with R.Plesser, A.Schwimmer, 
and especially S.Levit. This work was supported by a MINERVA fellowship.

\appendix

\section{Solution of fourth order equation determining
eigenvalue distribution of spinless particle}
\label{sp0ap}
The solutions of the equation (\ref{red4})
\begin{equation}
x^4 + (3N-6M^2 ) x^2 + (8M^3 - 6NM) x + 3NM^2 - 3M^4 + L/4 = 0
\end{equation}
are determined by the solutions of the corresponding cubic resolvent
equation \cite{bron}
\begin{equation}
y^3 + b_2 y^2 +b_1 y + b_0 = 0
\label{cubr}
\end{equation}
with
\begin{equation}
b_2 = 6N-12M^2 \ \ \ \ \ \ b_1 = 9N^2 - 48NM^2 + 48M^4 -L \ \ \ \ \ \ 
b_0 = -(8M^3 - 6NM)^2
\end{equation}
as
\begin{eqnarray}
x_1 = \frac{1}{2} (\sqrt{y_1 } + \sqrt{y_2 } - \sqrt{y_3 } )
& \ \ \ \ \ \ \ \ \ \ &
x_2 = \frac{1}{2} (\sqrt{y_1 } - \sqrt{y_2 } + \sqrt{y_3 } ) \nonumber \\
x_3 = \frac{1}{2} (-\sqrt{y_1 } + \sqrt{y_2 } + \sqrt{y_3 } )
& \ \ \ \ \ \ \ \ \ \ &
x_4 = \frac{1}{2} (-\sqrt{y_1 } - \sqrt{y_2 } - \sqrt{y_3 } )
\label{xpos}
\end{eqnarray}
where in addition to the signs given, the signs of the square roots
must be adjusted so as to guarantee
\begin{equation}
\sqrt{y_1 } \sqrt{y_2 } \sqrt{y_3 } = 8M^3 - 6NM
\label{sigc}
\end{equation}
In order to preserve continuity of the solutions in the parameters
$L,M,N$, this latter condition in practice is handled as follows: Choose the
square roots in the conventional way at some initial point in parameter space
where this is allowed. Points in parameter space where the
right hand side of (\ref{sigc}) changes sign coincide with one of
the solutions $y_i $ of (\ref{cubr}) touching zero; give the square
root of this solution an additional minus sign when crossing such a 
point to maintain (\ref{sigc}). Note that after two such sign changes
one may have switched between solutions in (\ref{xpos}).

The third order equation (\ref{cubr}) in turn can be simplified as follows.
Via the substitution
\begin{equation}
y=w-2N+4M^2
\end{equation}
it reduces to the normal form
\begin{equation}
w^3 + pw + q = 0
\label{cubn}
\end{equation}
with
\begin{equation}
p= -3N^2 -L \ \ \ \ \ \ \ \ q= -2N^3 + 2NL - 4M^2 L
\end{equation}
and the associated discriminant
\begin{equation}
d=\frac{p^3 }{27} + \frac{q^2 }{4} = 4L^2 (M^2 -3N/4)^2
+(4N^3 L + 2NL^2 ) (M^2 -3N/4) - \frac{N^2 L^2 }{12} - \frac{L^3 }{27}
\label{disc}
\end{equation}
The latter form will be useful below.

As a next step, it is necessary to pick out which of the solutions of the
original fourth order equation is the relevant one. The only place where
this question is easily answered is for $\xi \rightarrow \infty $. This
will be chosen as the starting point and the solution will then be
followed through parameter space using continuity. At
$\xi \rightarrow \infty $, using the definitions (\ref{pard}),
eq. (\ref{cubr}) simplifies to $y(y-4/a^4 )^2 = 0$ and thus two of the
solutions are $y_{1,2} = 4/a^4 $. For the third solution, one needs
to go to order $O(1/\xi^{2} )$. Writing $y_3 = s/\xi^2 $, and keeping
only the leading order in $1/\xi $ in (\ref{cubr}), one obtains
$s=(D-1)^2 /4 $. Note thus that (\ref{sigc}) is fulfilled to leading
order in $1/\xi $ with the conventional choice of the square roots.
Now, since the moment generating function $G_{H_0 } (\xi ) $ must vanish
as $\xi \rightarrow \infty $, and
\begin{equation}
G_{H_0 } (\xi ) = x(\xi ) - \frac{D-1}{4\xi } + \frac{2}{a^2 }
\end{equation}
(cf. (\ref{subs})), one easily picks out the solution $x_4 $ in
(\ref{xpos}) as the correct one for $\xi \rightarrow \infty $.

Now, to follow the solution through parameter space, it is necessary
to collect some facts about its behaviour. First, note that the constant
term in the cubic resolvent equation (\ref{cubr}) is negative. By
Vieta's theorem it thus follows that the product of its three solutions
is always positive. Therefore, the solutions of the equation may behave
in three qualitatively distinct ways:
\begin{enumerate}
\item Three positive real solutions: $\ \ d<0 \wedge b_2 <0 \wedge b_1 >0$.
\item One positive real solution, two complex conjugate solutions:
$\ \ d>0$.
\item One positive real solution, two negative real solutions:
$\ \ d<0 \wedge (b_2 > 0 \vee b_1 < 0)$.
\end{enumerate}
where in the characterization of alternatives 1.) and 3.), Descartes'
sign rule for the coefficients in (\ref{cubr}) has been used.

Further to this, one can show that, for $\xi >0$, changes of sign of
square roots according to condition (\ref{sigc}) can only occur in
regions of type 3.) in parameter space. The argument goes as follows:
The right hand side of (\ref{sigc}) being zero implies $4M^2 =3N$ (note
that $M>0$ for $\xi >0$). In these circumstances, the cubic resolvent
eq. (\ref{cubr}) simplifies to $y^3 -4M^2 y^2 -Ly =0$. Noting that 
$L>0 $ for $\xi >0$, this equation has one positive and one negative 
solution apart from the one which is zero. One is therefore indeed in a
region of type 3.) in parameter space; it is one of the two negative
solutions which touches zero and whose square root therefore acquires
a minus sign.

Below, slightly more than this will be needed, namely that when the solutions 
go from a region of type 2.) to a region of type 3.) and back to a region
of type 2.), either none or both the solutions which go from being complex
conjugate to negative and back to complex conjugate must have acquired a
minus sign for their square root. This is due to the fact that in regions
of type 2.), i.e. for $d>0$, the right hand side of (\ref{sigc}) is 
always positive. To see this, consider the converse proposition, namely
that $4M^2 -3N <0 $ implies $d<0$. First, note that
$4M^2 -3N < 0 \Rightarrow N>0$ and $4M^2 -3N > -3N$. Check that
at $4M^2 -3N =0$ and at $4M^2 -3N = -3N$, $d$ is manifestly negative;
furthermore, regarding (\ref{disc}) as a function of $4M^2 -3N$ and 
finding the lone extremum at $4M^2 -3N = -N-2N^3 /L$, one verifies that
also there, $d<0$, completing the argument.

Apart from sign changes in accordance with condition (\ref{sigc}), there is
one other possibility for a continuous switch between the different
solutions in (\ref{xpos}), namely that two of these solutions may
become degenerate (this is the generic case; higher degeneracies occur
only at exceptional points in parameter space). This happens precisely when
$d=0$, i.e. when there is a concomitant change of region. One can make this
more precise: When starting with the conventional choice of square roots in
(\ref{xpos}) in a region of type 1.), upon reaching a region of type 2.),
$x_4 $ always remains unique, only two of the other three solutions may
become degenerate. This remains true as long as one only alternates
between regions of type 1.) and type 2.). On the other hand, if one 
continues from a region of type 2.) to a region of type 3.) (and back),
$x_4 $ becomes degenerate with one of the other solutions; if, without
loss of generality, one denotes as $y_1 $ the solution of (\ref{cubr})
which has remained real and positive, then it is $x_4 $ and $x_3 $ which
become degenerate. This is the same switch which may also be effected
by condition (\ref{sigc}) as shown further above.

Armed with these properties, one can now argue that, despite all switches
of branch which may happen as one tracks the solution through parameter
space, it is only in regions of type 3.) that it may have a nonzero
imaginary part (which up to trivial factors gives the sought-after
eigenvalue distribution of the Hamiltonian (\ref{hs0})). As elucidated
above, at $\xi \rightarrow \infty $ one starts in a region of type 1.)
with the solution $x_4 $ in (\ref{xpos}). Now, changes of region occur
precisely where $d=0$. This latter condition is a fifth order equation
in $\xi $, so there can be at most five changes of region as $\xi $
is varied, i.e. the following sequences are possible\footnote{Note that
a direct transition between regions of type 1.) and type 3.) can only
occur if $b_0 =b_1 =0$ in (\ref{cubr}) for some value of $\xi $, which is
only possible at the exceptional point $k=0$.}:
\begin{enumerate}
\item[a.] type (1) $\rightarrow $ type (2) $\rightarrow $ type (1)
$\rightarrow $ type (2) $\rightarrow $ type (1) $\rightarrow $ type (2)
\item[b.] type (1) $\rightarrow $ type (2) $\rightarrow $ type (1)
$\rightarrow $ type (2) $\rightarrow $ type (3) $\rightarrow $ type (2)
\item[c.] type (1) $\rightarrow $ type (2) $\rightarrow $ type (3)
$\rightarrow $ type (2) $\rightarrow $ type (1) $\rightarrow $ type (2)
\item[d.] type (1) $\rightarrow $ type (2) $\rightarrow $ type (3)
$\rightarrow $ type (2) $\rightarrow $ type (3) $\rightarrow $ type (2)
\end{enumerate}
Of course, the sequences may be shorter if $d$ has less than five real
zeros as a function of $\xi $. Now, according to the above, in sequences
a.), b.), and d.) one always remains with solutions $x_4 $ and $x_3 $ (with
the convention that $y_1 $ is the solution which has remained real and 
positive throughout, otherwise $x_3 $ is replaced by one of the other
two solutions). However, both in $x_4 $ and $x_3 $ in regions of type
2.) the imaginary parts of $\sqrt{y_2 } $ and $\sqrt{y_3 } $ cancel;
thus one indeed only has a nonzero imaginary part of the solution in regions
of type 3.). The remaining sequence c.) is more complicated. There, one may
emerge in the second region of type 1.) with the solution $x_3 $ and thus,
going to the last region of type 2.), switch to a solution in which the
imaginary parts do not cancel anymore. Here, this last scenario will not
be analyzed in detail. In practice, it suffices to note that the region
of type 3.) already saturates the normalization of the eigenvalue distribution
and that therefore there can not be any additional nonzero contribution
from the last region of type 2.).

Summing up, the imaginary part of $G_{H_0 } (\xi ) $ is only nonzero
in regions of type 3.), i.e. formally one can write
\begin{equation}
\mbox{Im} \, G_{H_0 } (\xi ) = -\frac{1}{2} \theta (-d)
(1-\theta (-b_2 ) \theta (b_1 )) \mbox{Im} \, (\pm \sqrt{y_1 }
\pm \sqrt{y_2 } \pm \sqrt{y_3 } )
\label{imgt}
\end{equation}
where $\theta $ denotes the step function. The remaining ambiguity will be
settled presently using condition (\ref{sigc}) and the positivity of the
eigenvalue distribution. To this end, one must specify more explicitly the
solutions $w_i $ of (\ref{cubn}). In the case
$d<0$, which is the only one of interest here, (\ref{cubn}) is
solved by
\begin{equation}
w_1 = 2\mbox{Re} \, (u) \ \ \ \ \ \ 
w_2 = 2\mbox{Re} \, (u e^{2\pi i/3} ) \ \ \ \ \ \
w_3 = 2\mbox{Re} \, (u e^{-2\pi i/3} )
\label{cnsol}
\end{equation}
with
\begin{equation}
u=\left( -\frac{q}{2} + \sqrt{d} \right)^{1/3}
\label{udef}
\end{equation}
Choosing the roots in (\ref{udef}) in the conventional way (i.e.
$\sqrt{-1} = i$, any other choice merely permutes the $w_i $), one has
$w_1 \ge w_3 \ge w_2 $, and consequently $y_1 \ge 0 \ge y_3 \ge y_2 $,
i.e. $\mbox{Im} \, \sqrt{y_1 } =0$, the convention which was already
used above. Furthermore, upon entering a region of type 3.) from a
region of type 2.), $\sqrt{y_2 } $ and $\sqrt{y_3 } $ must be chosen to 
lie in different half-planes (upper or lower) in accordance with 
continuity and condition (\ref{sigc}), the right hand side of which is
positive at the interface, $4M^3 -3NM >0$. Only when within the region
of type 3.), $4M^3 -3NM$ becomes negative, the root which touches zero
acquires an additional minus sign; this is $\sqrt{y_3 } $ according to
the above conventions. Therefore, one can now specify 
$\mbox{Im} \, G_{H_0 } $ up to an overall sign,
\begin{equation}
\mbox{Im} \, G_{H_0 } (\xi ) = \pm \frac{1}{2} \theta (-d)
(1-\theta (-b_2 ) \theta (b_1 )) \left[
\sqrt{|y_2 |} + (1-2\theta(4M^3 -3NM)) \sqrt{|y_3 |} \right]
\end{equation}
and this last ambiguity is resolved using the positivity of the
eigenvalue distribution, which thus reads
\begin{eqnarray}
\rho_{H_0 } (\xi ) &=& -\frac{1}{\pi } \mbox{Im} \, G_{H_0 } (\xi ) \\
&=& \frac{1}{2\pi } \theta (-d) (1-\theta (-b_2 ) \theta (b_1 ) )
\left| \sqrt{ | y_2 | } + (1-2\theta (4M^3 -3NM )) \sqrt{ | y_3 | } \right|
\end{eqnarray}

\section{Eigenvalue distribution in the polar representation}
In the polar representation, expectation values
\begin{eqnarray}
\langle O \rangle &=& \frac{1}{Z} \int d^N \eta \, J(\vec{\eta } )
O(\vec{\eta } )
\exp ( -N \beta \vec{\eta }^{2} ) \\
&=& \int d^N \eta \, O(\vec{\eta } ) \exp \left(
-N\beta \sum_{i=1}^{N} \eta_{i}^{2} + 
\sum_{i<j} \ln (\eta_{i}^{2} -\eta_{j}^{2} )^2
+\sum_{i=1}^{N} \ln \eta_{i} \right)
\label{polex}
\end{eqnarray}
are dominated by the saddle points of the exponent for large $N$.
Note that the last term in the exponent in eq. (\ref{polex}) is only
of order $O(N)$ as opposed to $O(N^2 )$ for the other two terms.
Thus, going to a continuous eigenvalue distribution\footnote{The
subscript makes explicit that $\rho_{B} $ is the eigenvalue distribution
of the matrix $B$ with eigenvalues $\eta_{i} $. Below, also
$\rho_{B^2 } (x) = \rho_{B} (\sqrt{x} )/2\sqrt{x} $ will be considered.}
$\rho_{B} (\eta ) $ for $N\rightarrow \infty $, the saddle point
distribution is determined by
\begin{eqnarray}
0 &=& \frac{\delta }{\delta \rho_{B} (\eta ) } \left[
\frac{1}{2} \int_{c}^{b} dx \, dy \, \rho_{B} (x) \rho_{B} (y)
\ln (x^2 -y^2 )^2 - \beta \int_{c}^{b} dx \, x^2 \rho_{B} (x) \right]
\label{paction} \\
&=& \int_{c}^{b} dx \, \rho_{B} (x) \ln (x^2 -\eta^{2} )^2 -\beta \eta^{2}
\end{eqnarray}
or, taking the derivative with respect to $\eta $ and dividing by $\eta $,
\begin{equation}
0=2\int_{c}^{b} dx \, \frac{\rho_{B} (x) }{x^2 -\eta^{2} } + \beta
\end{equation}
Here, the (as yet undetermined) edges of the support of $\rho_{B} $
have been made explicit. Note that $c$ and $b$ are positive because
the eigenvalues $\eta_{i} $ are by definition positive.
Substituting $x^2 =s$ and $\eta^{2} =t$, one obtains
\begin{equation}
-\beta = 2\int_{c^2 }^{b^2 } ds \, \frac{\rho_{B^2 } (s) }{s-t}
\label{sqsadd}
\end{equation}
After defining
\begin{equation}
\tilde{\rho }_{B^2 } (s) = \rho_{B^2 } (s) \sqrt{\frac{s-c^2 }{b^2 -s} }
\end{equation}
one can use the inversion formulae on a finite interval \cite{akh}
\begin{equation}
g(x) = \frac{1}{\pi } \int_{c^2 }^{b^2 } dy \,
\sqrt{\frac{b^2 -y}{y-c^2 } } \frac{f(y)}{y-x} \ \ \ \ \ \ \ \ 
f(x) = \frac{1}{\pi } \int_{c^2 }^{b^2 } dy \,
\sqrt{\frac{y-c^2 }{b^2 -y} } \frac{g(y)}{x-y}
\end{equation}
to obtain
\begin{equation}
\tilde{\rho }_{B^2 } (s) = \frac{\beta }{2\pi } \ \ \ \ \ \ 
\rho_{B^2 } (s) = \frac{\beta }{2\pi } \sqrt{\frac{b^2 -s}{s-c^2 } }
\ \ \ \ \ \ 
\rho_{B} (\eta ) = \frac{\beta \eta }{\pi } 
\sqrt{\frac{b^2 -\eta^{2} }{\eta^{2} -c^2 } }
\label{bdis}
\end{equation}
Now, the condition that $\rho_{B^2 } $ must be normalized to one can be
used to determine $b$,
\begin{equation}
b^2 = \frac{4}{\beta } + c^2
\end{equation}
On the other hand, the lower bound $c$ is not determined by any auxiliary
condition and should also be varied when determining the saddle point
distribution. In other words, up to now, not the full space of possible
$\rho_{B} $ has been explored, but only the subspace with fixed, albeit
arbitrary, lower bound $c$. To determine $c$, one inserts the form
(\ref{bdis}) into the ``action'' in the square brackets in
(\ref{paction}) and now varies with respect to $c$. The first term
in the action can be made manifestly independent of $c$ by again
substituting as in eq. (\ref{sqsadd}) and a shift in the integration
variables; thus one very easily obtains $c=0$, as one would expect.
Finally, inserting the radius of the original semicircular distributions
of the gauge fields, $a^2 =2/\beta $, one has
\begin{equation}
\rho_{B^2 } (s) = \frac{1}{\pi a^2 } \sqrt{\frac{2a^2 -s }{s} }
\end{equation}
on the interval $[0,2a^2 ] $ for the eigenvalue distribution of $B^2 $.

\section{Solution of fourth order equation determining
eigenvalue distribution of particle with spin}
The solutions of the equation (\ref{spin4})
\begin{equation}
x^4 + c_2 x^2 + c_1 x + c_0 = 0
\label{spin44}
\end{equation}
with
\begin{eqnarray}
c_2 &=& -\frac{8}{3} (2L^2 + 4KL + 6K^2 + 1) \\
c_1 &=& -\frac{64}{27} (2L^3 + 12KL^2 + 18K^2 L + 3L -9K ) \\
c_0 &=& -\frac{16}{27} (32KL^3 + 48K^2 L^2 +8L^2 - 48KL +9)
\end{eqnarray}
are determined by the solutions of the corresponding cubic resolvent
equation
\begin{equation}
y^3 + b_2 y^2 +b_1 y + b_0 = 0
\label{cubrsp}
\end{equation}
with
\begin{eqnarray}
b_2 &=& -\frac{16}{3} (1+2K^2 +L^2 + (2K+L)^2 ) < 0 \\
b_1 &=& \frac{256}{27} (3+4L^2 +(3K-L)^2 + 24L^2 (K+L/3)^2 + (3K+L)^4 /3 )
>0 \\
b_0 &=& -\frac{4096}{729} (2L^3 + 12KL^2 + 18K^2 L +3L -9K)^2 < 0
\end{eqnarray}
as in eq. (\ref{xpos}). Additionally, the signs of the square roots
of the $y_i $ in (\ref{xpos}) must be adjusted so as to guarantee
\begin{equation}
\sqrt{y_1 } \sqrt{y_2 } \sqrt{y_3 } = c_1 =
-\frac{64}{27} (2L^3 + 12KL^2 + 18K^2 L + 3L -9K )
\label{sigc4}
\end{equation}
The third order equation (\ref{cubrsp}) in turn can be simplified
by the substitution $y=w-b_2 /3 $ to
\begin{equation}
w^3 + pw + q = 0
\label{cubnsp}
\end{equation}
with
\begin{eqnarray}
p &=& \frac{256}{27} (2-3K^2 -9K^4 -14KL -12K^3 L +L^2 +2K^2 L^2 
+4KL^3 -L^4 ) \\
q &=& \frac{4096}{729} (7-36K^2 +27K^4 +54K^6 +48KL +144K^3 L +108K^5 L
+12L^2 \nonumber \\ & & \ \ \ \ \ \ \ \ \ \ 
+66K^2 L^2 +18K^4 L^2 -48KL^3 -56K^3 L^3 +3L^4 -6K^2 L^4
+12KL^5 -2L^6 )
\end{eqnarray}
and the associated discriminant
\begin{eqnarray}
d &=& \frac{p^3 }{27} + \frac{q^2 }{4} \\
&=& \frac{4194304}{19683} (3-24K^2 +54K^4 -81K^8 +8L^2 +260 K^2 L^2
+384 K^4 L^2 \nonumber \\ & & \ \ \ \ \ \ \ \ \ \ \ \ \ \ \ 
+ 180 K^6 L^2 + 6L^4 -128 K^2 L^4 -118 K^4 L^4 + 20 K^2 L^6 -L^8 )
\label{disc4}
\end{eqnarray}
The analysis of these equations turns out to be considerably simpler
than for the equations describing the case of the spinless particle.
There is only one property which is not immediately obvious, namely
that the discriminant (\ref{disc4}) has at most two zeros as a function
of $L$ for $L\in [0,\infty ]$ (remember that the whole
problem is symmetrical about $L=0$ and it is therefore sufficient
to consider positive $L$). One proves this by regarding $d$ as a
polynomial in $L^2 $ and constructing the Sturm chain
\cite{bron} of $d(L^2 )$ at $L^2 =0$ and for
$L^2 \rightarrow \infty $. The absolute difference between the
number of sign changes occuring in the one or the other chain gives the
number of zeros of $d(L^2 )$. One can check that this is less
than or equal to two for any $K$, where it turns out to be advantageous
to distinguish between the cases $K^2 \in [0,1/60], K^2 \in [1/60,1/3]$, and
$K^2 \in [1/3,\infty ]$.

Now, just as in the case of the spinless particle 
(cf. Appendix \ref{sp0ap}), the solutions of the
cubic resolvent equation (\ref{cubrsp}) can in principle display three
qualitatively different types of behaviour, partitioning the space of
parameters $K$ and $L$ into regions corresponding to the different types.
Here, however, there are no regions of type 3.), since one always has
$b_2 <0 $ and $b_1 >0$. Therefore, the support of the eigenvalue
distribution of the Pauli Hamiltonian must come from a region of type
2.) in which the imaginary parts of the two complex conjugate solutions
do not cancel. Moreover, due to the fact that $d$ has at most two
zeros as $L$ is varied over all positive values, and
$d<0 $ for $L \rightarrow \infty $, there is only one unique
region of type 2.) for $L>0$ (and its mirror image for
$L<0$). Denoting without loss of generality as $y_1 $ and $y_2 $
the two complex conjugate solutions and as $y_3 $ the real
positive solution, the form of the relevant solution 
$x \equiv aG_{Q_{\lambda } } (0) -4L/3$ of (\ref{spin44}) on the region
of type 2.) is therefore already almost uniquely determined:
\begin{equation}
\mbox{Im} \, x = -\frac{1}{2} | i (\sqrt{y_1 } -\sqrt{y_2 } ) |
\ \ \ \ \ \ \ \ \ \ 
\mbox{Re} \, x = \pm \frac{1}{2} \sqrt{y_3 }
\label{choisp}
\end{equation}
where in Im$x$ it has additionally been used that due to the positivity
of the eigenvalue distribution, $\mbox{Im} \, x = a \mbox{Im} \,
G_{Q_{\lambda } } (0)$ must be chosen such that it is negative
(cf. eq. (\ref{reimsp})). The ambiguity in Re$x$ on the other hand is
resolved with the help of condition (\ref{sigc4}). When the right hand side
of (\ref{sigc4}) is positive, it is consistent to choose the square roots
$\sqrt{y_i } $ in the conventional way and (\ref{choisp}) can then be
identified with one of the two solutions $x_2 $ or $x_3 $ in (\ref{xpos}),
i.e. $\sqrt{y_3 } $ then comes with the positive sign in (\ref{choisp}).
On the other hand, if a sign change occurs in condition (\ref{sigc4}) within
the region of type 2.), then it can only be the real positive solution
$y_3 $ which touches zero, i.e. it is $\sqrt{y_3 } $ which acquires an
additional minus sign. In other words, one has
\begin{equation}
\mbox{Re} \, x = \frac{1}{2} (2\theta 
(9K - 2L^3 - 12KL^2 - 18K^2 L - 3L ) -1) \sqrt{y_3 }
\end{equation}
It remains to give the solutions $w_i $ of equation 
(\ref{cubnsp}) explicitly. In a region of type 2.), i.e. for $d>0$, they are
\begin{equation}
w_1 = e^{2\pi i/3} u + e^{-2\pi i/3} u^{\prime } \ \ \ \ \ \ 
w_2 = e^{-2\pi i/3} u + e^{2\pi i/3} u^{\prime } \ \ \ \ \ \
w_3 = u+u^{\prime }
\end{equation}
where
\begin{equation}
u= \frac{p}{|p|} \left| -\frac{q}{2} + \sqrt{d} \right|^{1/3}
\ \ \ \ \ \ \ \ \ \ 
u^{\prime } = -\frac{|p|}{3} \left| -\frac{q}{2} + \sqrt{d} \right|^{-1/3}
\end{equation}
This fixes the solutions $y_i = w_i - b_2 /3$ of the cubic resolvent
equation (\ref{cubrsp}), in terms of which the solution for 
$G_{Q_{\lambda } } (0)$ is now finally given as
\begin{eqnarray}
\mbox{Im} \, G_{Q_{\lambda } } (0) &=& 
-\frac{1}{2a} \theta (d) | i (\sqrt{y_1 } -\sqrt{y_2 } ) | \\
\mbox{Re} \, G_{Q_{\lambda } } (0) &=& \frac{1}{2a} (2\theta
(9K - 2L^3 - 12KL^2 - 18K^2 L - 3L ) -1) \sqrt{y_3 }
+\frac{4L}{3a} \ \ \ \mbox{for} \ d>0
\end{eqnarray}

\begin{figure}
\caption{$\rho_{\protect\sqrt{H_{0} } } (\lambda )$ for different momenta
$k$, in $D=3$ dimensions.}
\label{f1}
\end{figure}

\begin{figure}
\caption{$\rho_{\protect\sqrt{H_{0} } } (\lambda )$ for different dimensions
$D$, with momentum $k=0$.}
\label{f2}
\end{figure}

\begin{figure}
\caption{$\rho_{\protect\sqrt{H_{0} } } (\lambda )$ for different dimensions
$D$, with momentum $k=0$, rescaled corresponding to keeping a fixed norm
of the vector field as $D$ is varied (see text).}
\label{f3}
\end{figure}

\begin{figure}
\caption{Left-hand graph: Loci of solutions of det$(Q_{\lambda } -\mu )=0$ 
for a generic realisation of the gauge matrices $A_i $. Changing the
momentum $k$, which enters the above equation as a parameter, merely
amounts to shifting the origin of the $\lambda -\mu $-plane in the
direction of the line $\lambda =-\mu /2$ (see text). Right-hand graph: 
Non-monotonous behaviour of trajectories in the $\lambda -\mu $-plane
is impossible (see text). Dotted lines serve merely to guide the eye.}
\label{traj}
\end{figure}

\begin{figure}
\caption{$\rho_{\protect\sqrt{H_{S} } } (\lambda )$ for different momenta
$k$. The reader is reminded that $H_S $ was a $2N\times 2N$ matrix, and that
that therefore the eigenvalue distribution of $\protect\sqrt{H_{S} } $
as constructed in the text is normalised to two.}
\label{ff1}
\end{figure}

\begin{figure}
\caption{$\rho_{\protect\sqrt{H_{0} } } $ (solid lines) and
$\rho_{\protect\sqrt{H_{S} } } /2 $ (dashed lines) in three dimensions
for (from left to right) $k/a = 0,\frac{1}{2},1,2,5$ (note the shift in
the $\lambda $-axis for the last two plots).}
\label{fc1}
\end{figure}

\end{document}